\pdfoutput=1
\documentclass[preprint]{aastex61}
\usepackage{natbib}
\usepackage{rotating}
\usepackage{upgreek}
\usepackage{amsmath}
\usepackage{amssymb}
\usepackage{graphicx}
\usepackage{color}

\newcommand{\ms}{m s$^{-1}$ }

\begin{document}

\title{A Low-cost Environmental Control System for Precise Radial Velocity Spectrometers}

%\author{David H. Sliski\altaffilmark{1}, Cullen H. Blake\altaffilmark{1}, Samuel Halverson\altaffilmark{1},\altaffilmark{2}}
%\affiliation{\altaffilmark{1}University of Pennsylvania, Department of Physics and Astronomy, 209 S 33rd St, Philadelphia, PA 19104, USA}
%\affiliation{\altaffilmark{2}Sagan Fellow}

\author{David H. Sliski}
\affiliation{University of Pennsylvania, Department of Physics and Astronomy, 209 S 33rd St, Philadelphia, PA 19104, USA}
\author{Cullen H. Blake}
\affiliation{University of Pennsylvania, Department of Physics and Astronomy, 209 S 33rd St, Philadelphia, PA 19104, USA}
\author{Samuel Halverson}
\affiliation{University of Pennsylvania, Department of Physics and Astronomy, 209 S 33rd St, Philadelphia, PA 19104, USA}
\affiliation{Sagan Fellow}

\begin{abstract}
We present an Environmental Control System (ECS) designed to achieve milliKelvin (mK) level temperature stability for small-scale astronomical instruments. This ECS is inexpensive and is primarily built from commercially available components. The primary application for our ECS is the high-precision Doppler spectrometer MINERVA-Red, where the thermal variations of the optical components within the instrument represent a major source of systematic error. We demonstrate $\pm 2$~mK temperature stability within a 0.5 m$^{3}$ Thermal Enclosure using resistive heaters in conjunction with a commercially available PID controller and off-the-shelf thermal sensors. The enclosure is maintained above ambient temperature, enabling rapid cooling through heat dissipation into the surrounding environment. We demonstrate peak-to-valley (PV) temperature stability of better than 5~mK within the MINERVA-Red vacuum chamber, which is located inside the Thermal Enclosure, despite large temperature swings in the ambient laboratory environment. During periods of stable laboratory conditions, the PV variations within the vacuum chamber are less than 3~mK. This temperature stability is comparable to the best stability demonstrated for Doppler spectrometers currently achieving 1 m s$^{-1}$ radial velocity precision. We discuss the challenges of using commercially available thermoelectrically cooled CCD cameras in a temperature-stabilized environment, and demonstrate that the effects of variable heat output from the CCD camera body can be mitigated using PID-controlled chilled water systems. The ECS presented here could potentially provide the stable operating environment required for future compact, ``astro-photonic'' precise radial velocity (PRV) spectrometers to achieve high Doppler measurement precision with a modest budget.

\end{abstract}

\section{Introduction}

The measurement of the reflex motion of an exoplanet host star through the detection of minute Doppler shifts in stellar spectral features is crucial to ongoing efforts to discover exoplanets and measure their bulk properties. Moreover, in transiting systems Doppler measurements allow for a direct measurement of planetary mass. However, these measurements are a significant technical challenge for modern spectrometers. For example, the reflex motion of the Sun resulting from Earth's orbit requires that the relative positions of stellar spectral lines are measured to a fraction of a \textit{femtometer} ($\sim$10 c{\ms} velocity semi-amplitude). Detecting exoplanets using the Doppler method requires exquisite calibration of the wavelength scale of the spectrometer and careful attention to instrumental effects that can masquerade as Doppler signals. Many of these instrumental effects can be mitigated by placing the spectrometer in a vacuum environment and keeping the temperature of instrument components very stable \citep{NEIDECS}. A vacuum environment is necessary to eliminate the wavelength shifts resulting from variations in the index of refraction of air, and temperature stability ensures that thermally-induced changes in optical and optomechanical components are kept to a minimum. This ``stabilized'' Doppler spectrometer approach was first implemented with the HARPS instrument \citep{HARPS}, and subsequently many others, to make stellar radial velocity measurements at the m s$^{-1}$ level. This approach will be improved with upcoming extremely precise Doppler spectrometers like NEID \citep{NEID}  and ESPRESSO \citep{ESPRESSO} having 10 cm s$^{-1}$ precision goals. 

Given upcoming transit surveys targeting bright stars, such as TESS \citep{TESS} and PLATO \citep{PLATO}, sub-meter class telescopes will play an increasingly important role in exoplanet characterization efforts. Since the physical size, and therefore cost, of a spectrometer at fixed resolution and astronomical site quality is proportional to telescope diameter, Doppler measurements with a precision comparable to the current state of the art become possible with a modest budget by incorporating off-the-shelf components into the instrument. This rich scientific niche is being pursued by several current and upcoming projects, such as MINERVA \citep{MINERVA}, MINERVA-Red \citep{MINERVARed}, MICRONERVA \citep{MICRONERVA}, the Dharma Planet survey \citep{Dharma} and RHEA \citep{RHEA}. Similarly, astrophotonic technologies and extreme adaptive optics systems enable efficient coupling of light from a large telescope into single-mode fiber, making it possible to also build small spectrometers for the world's largest telescopes like the iLocator spectrograph on the LBT \citep{iLocator} and the photonic spectrographs being tested with the Subaru Coronagraphic Extreme Adaptive Optics (SCExAO, \citealt{Subaru, NEM}) system on the Subaru telescope.

We describe an inexpensive ECS designed to provide a thermally stable environment for ``table top" Doppler spectrometers. This system combines many commercially available parts to achieve~mK temperature stability. In Section 2 we briefly describe MINERVA-Red, the project for which this ECS was designed. In Section 3 we outline the minimum thermal requirements to achieve 1 m s$^{-1}$ instrumental Doppler precision. In Section 4 we describe the primary components of the ECS, as well as potential thermal links between a vacuum chamber inside the Thermal Enclosure and the outside world. In Section 5 we summarize the performance of the ECS over a one-month test, and in Section 6 we present conclusions and directions for future improvement of our ECS.

\section{MINERVA-Red}

The ECS described here has been designed for a cross-dispersed Echelle spectrometer being built for the MINERVA-Red project. The goal of this project is intensive Doppler monitoring of nearby low-mass stars. The MINERVA-Red spectrometer is optimized for red optical wavelengths, where these low-mass stars are brightest and their spectra are rich with spectral lines (e.g. \citealt{mdwarf}). The spectrometer covers a modest wavelength range of 820 to 920 nm using a red-sensitive, deep depletion CCD detector. The spectrometer operates inside a round vacuum chamber that is 75 centimeter in diameter and 30 cm in height. This small size is made possible by the fact that the instrument is designed for a 70 cm telescope and will be coupled to the telescope via a single-mode fiber. Wavelength calibration will be achieved with UNe lamps, though a custom Fabry-Perot Etalon is also being explored \citep{SAMFIBER}.

\section{Thermal Requirements for Intrinsic Stability}
A comprehensive error budget for precise Doppler measurements includes many sources of systematic errors, including those related to modal noise in optical fibers, correction of telluric absorption, and the stability of the wavelength calibration source \citep{SAMNEID}. However, a leading source of measurement error for an instrument that is not in a controlled environment relates to the thermally-induced expansion, contraction, or deformation of instrument components such as the diffraction grating and optical mounts. While the exact response of a specific instrument to a given temperature fluctuation requires detailed simulations of the entire optomechanical system, a first approximation of temperature stability requirements can be derived from calculating the impact of expansion of the primary dispersive element on the location of monochromatic light on the instrument focal plane.  Most modern Doppler spectrometers employ an Echelle grating and cross disperser to simultaneously achieve high spectral resolution over a wide wavelength range. The MINERVA-Red spectrometer employs an R2 Echelle grating similar to that found in many Doppler spectrometers, with key parameters of the grating supplied by Newport shown in Table 1. 

\begin{deluxetable}{ c c }[h]
\tablecaption{MINERVA-Red Echelle Grating Parameters}
 \startdata
 \\
 Wavelength Range & 820-920 nm \\
 n (Echelle order) & 62-68 \\ 
 CTE (Zerodur) & $<10^{-7}$ K$^{-1}$  \\  
 d ($\sigma$) &  31600 nm spacing (31.6 lines per mm) \\ 
 Blaze Angle  & 63$^\circ$ \\
 Grating Size & 50 x 100 mm\\
 \enddata
\end{deluxetable}

As the temperature of the grating changes, the substrate expands or contracts, altering the spacing of the grating grooves. Assuming that changes in the illumination angle of the grating are negligible, then this change in the groove spacing, $d$, effectively changes the central wavelength of each diffraction order as recorded on the focal plane. The MINERVA-Red spectrometer is in a quasi-Littrow configuration where the angle of illumination, $\alpha$, is approximately equal to the angle of diffraction, $\beta$, which is approximately equal to the blaze angle of the grating. In this case, the grating equation becomes
\begin{equation}
n\lambda = d[\sin(\alpha) +\sin(\beta)] \approx 2d\sin(\alpha)
\end{equation}
The change in the central wavelength of each diffraction order, $\Delta \lambda$, is then
\begin{equation}
\Delta\lambda = \frac{2\sin(\alpha)\Delta d}{n} 
\end{equation}
The change in distance between the grooves, $\Delta d$, is a function of the grating material properties. The MINERVA-Red Echelle grating substrate is Zerodur, a very low Coefficient of Thermal Expansion (CTE) material. For Class 2 Zerodur, $ |CTE|<10^{-7}/K $. For a temperature change $\Delta T$ we can calculate the change in groove spacing 
\begin{equation}
\Delta d = d \times \mathrm{CTE} \times \Delta\mathrm{T}
\end{equation}
and the induced wavelength shift is
\begin{equation}
\Delta\lambda = \frac{2\sin(\alpha)\times d\times \mathrm{CTE} \times \Delta\mathrm{T}}{n}=\lambda\times \mathrm{CTE}\times \Delta T
\end{equation}
For $\Delta T=0.01$K, $\Delta\lambda$ is $8.94 \times 10^{-7}$ nm at the middle of the MINERVA-Red wavelength range at 850 nm in order $n=65$. This is equivalent to a velocity shift $\Delta v = c \times{\frac{\Delta\lambda}{\lambda}} = 0.306$ m s$^{-1}$. Since a key goal of the MINERVA-Red project is 1 m s$^{-1}$ velocity measurements for bright, low-mass stars, we have designed our ECS to achieve long-term temperature stability better than $\pm 10$~mK. We expect that the fraction of the total radial velocity error budget that is attributable to thermal effects is comparable to the intrinsic error in our wavelength reference. This means that thermal stability represents one of the larger terms in our overall instrumental error budget.

\begin{figure}[htp]
\centering
\includegraphics[width=12cm, trim={1cm 6cm 2cm 7cm},clip]{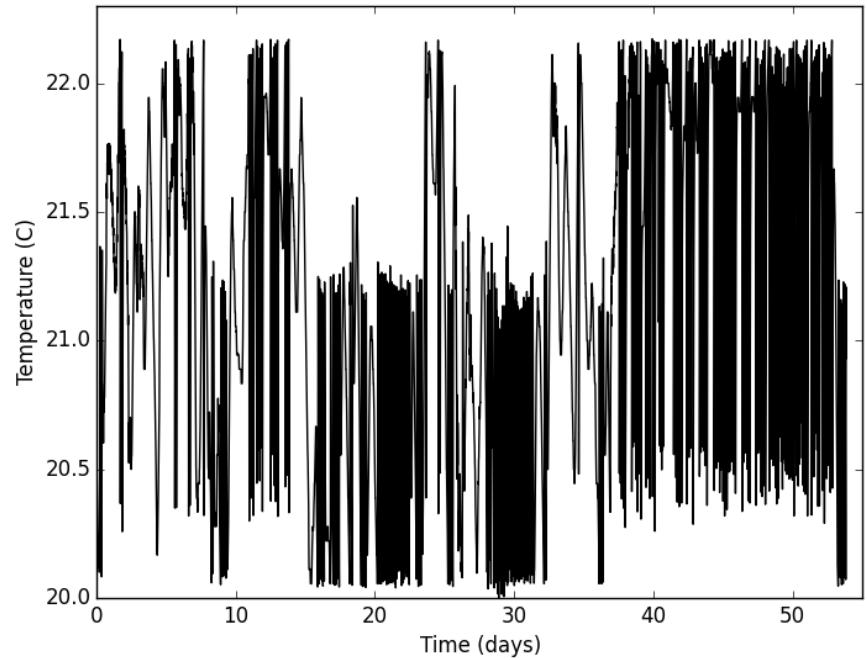}
\caption{Temperature data for the MINERVA spectrograph room located at the Fred Lawrence Whipple Observatory on Mt. Hopkins in Amado, Arizona. Data is recorded every 30 seconds, but shown here averaged in 15 minute bins.}
\label{fig:mthopkinstempdata}
\end{figure}

The thermal stability of any spectrometer will be inevitably limited by the stability of the surrounding environment. A major advantage of fiber-coupled spectrometer is that it is possible to place the instrument in an environment that is much more benign than the telescope dome. Anecdotal evidence suggests that many laboratory settings provide $\pm 1^{\circ}$C stability. The MINERVA facility at Fred Lawrence Whipple Observatory at Mt. Hopkins, Arizona, includes a temperature controlled room for both the MINERVA and MINERVA-Red spectrometers. We find that this room, which was specifically designed to house Doppler spectrometers, undergoes long-term temperature changes of up to $\pm 0.75^{\circ}$C as seen in Figure \ref{fig:mthopkinstempdata}. We have designed and tested our ECS in a laboratory environment providing similar temperature stability.
 
\begin{figure}[htp]
\centering
\includegraphics[width=10cm, angle = 270]{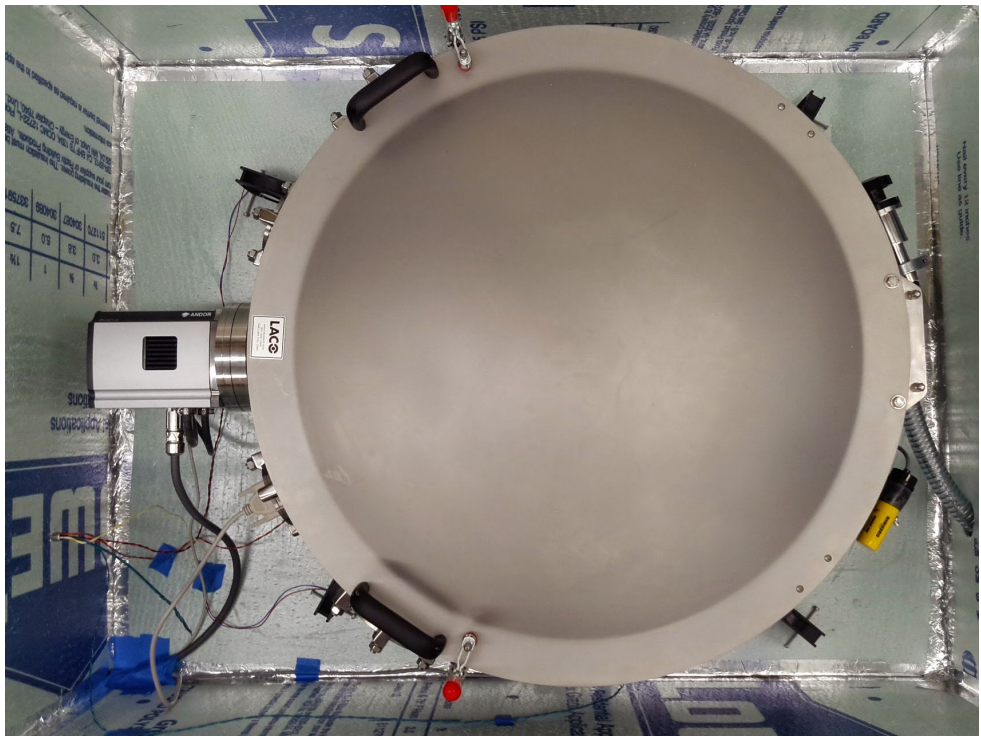}
\caption{MINERVA-Red vacuum chamber inside the Thermal Enclosure. The chamber, which is made of Aluminum, is 75 cm in diameter. The heater units are not shown here. }
\label{fig:chamber}
\end{figure}

\section{The Environmental Control System }
Our approach to minimizing $\Delta$T relies on precise regulation of the air temperature in the immediate vicinity of the MINERVA-Red instrument, relying on multiple layers of insulation to damp variations in the ambient environment, and minimizing any thermally conductive paths between the outside environment and the instrument optical components. We operate the instrument above ambient temperature, allowing for rapid cooling by dissipating heat into the outside environment. This approach is similar to the approach taken for the ECS for NEID, which achieves long-term stability of a fraction of a~mK \citep{NEIDECS}. Throughout the design of the ECS, we have selected low-cost, commercially available components whenever possible. 

The outer layer of the ECS is a box, called the Thermal Enclosure, that is made of polyisocyanurate, a common form of rigid insulation found at most construction stores. We employed 2" thick foam with an advertised R-value of R10 (R5 per inch of thickness). R-value is a measure of thermal resistance. The R$\#$ refers to imperial units of degrees Fahrenheit square feet hours per Btu (ft$^2$~$^{\circ}$F h Btu$^{-1}$). This unit is common when evaluating insulation products at large constructions stores in the United States. The SI unit for thermal resistance is Kelvin square meters per watt (K m$^2$W$^{-1}$). Internationally, this unit is described as an RSI value. In this case R10 is equivalent to an RSI of 1.76. The dimensions of the box were selected to accommodate the MINERVA-Red vacuum chamber while providing at least 1" of clearance around all sides of the chamber. The inner dimension of the box measures 36" by 56" by 16" and the box was assembled using 2" HVAC foil tape at joints and corners both inside and out. The box has eight holes, three for insulating feet which mount the chamber to the large optical bench below, one for a vacuum hose, two for the liquid supply for the CCD camera, and two for cables. Each of these holes was then back filled with polyurethane spray foam insulation to ensure a tight seal. The Thermal Enclosure is mounted to an optical table using PEEK offsets so that there is an air gap between the foam board and the optical table. PEEK is a machinable plastic that has very low thermal conductivity ($\approx 0.25$~W m$^{-1}$K$^{-1}$), about 800 times less than Aluminum. The Thermal Enclosure with the MINERVA-Red vacuum chamber inside is shown in Figure \ref{fig:chamber}.

\begin{figure}[htp]
\centering
\includegraphics[width=9cm]{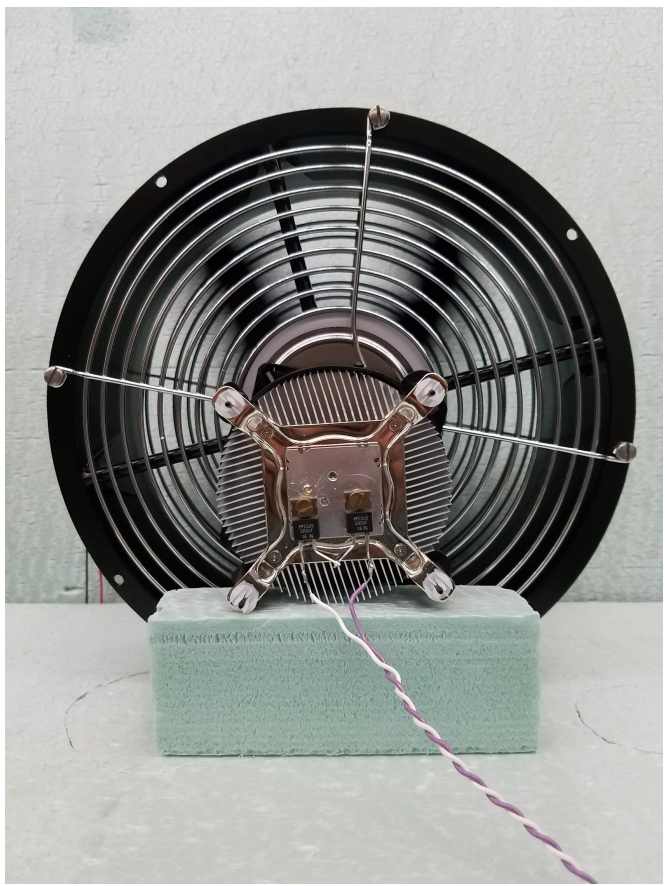}
\caption{Heater unit with 30 $\Omega$ resistive heaters attached to a CPU heatsink and 10" fan. Each heater unit can provide up to 50 watts of heat and is controlled by the SRS PTC10 temperature controller.}
\label{fig:heater}
\end{figure}

\begin{figure}[htp]
\centering
\includegraphics[width=13cm, angle = 270]{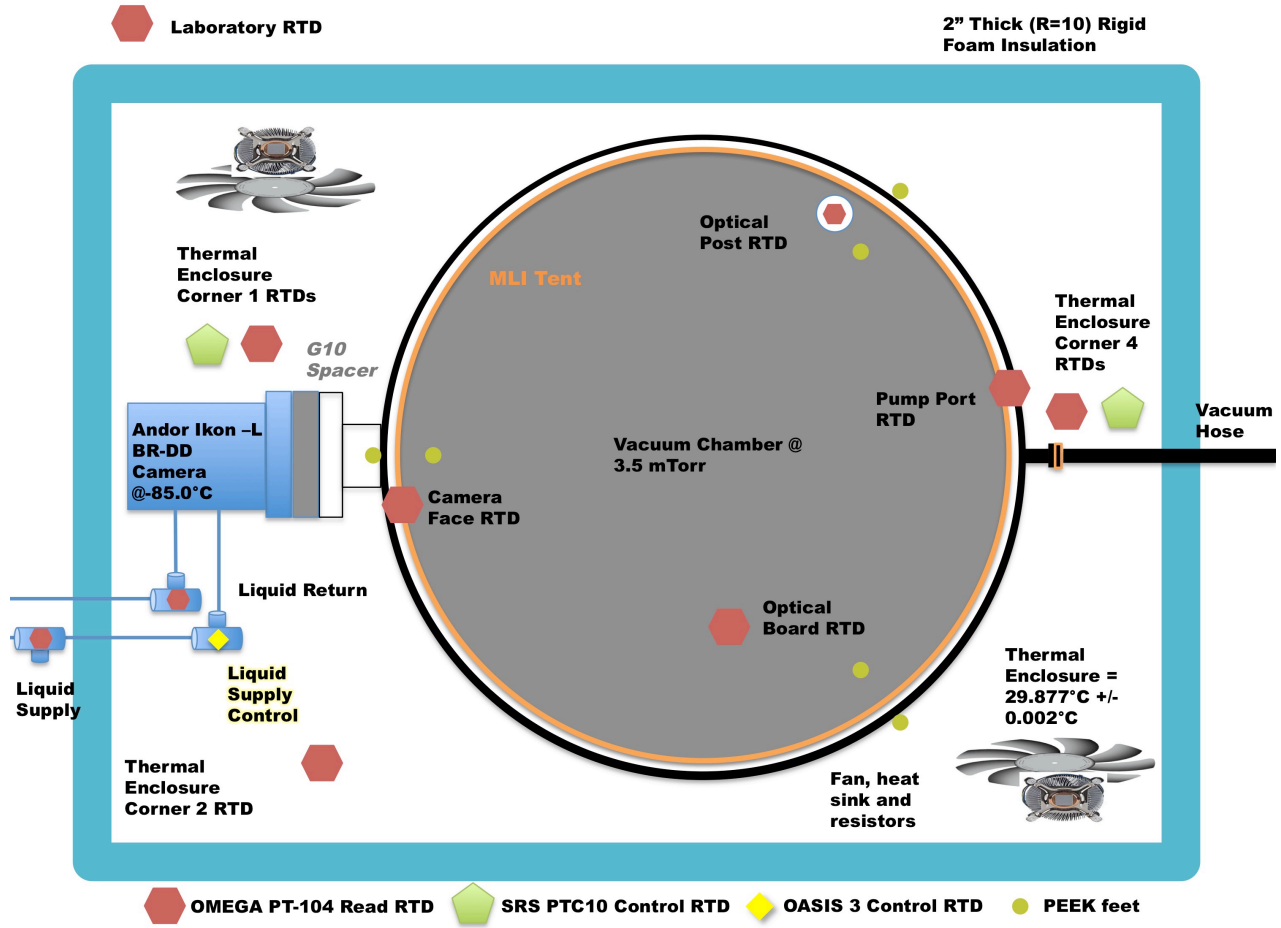}
\caption{Schematic of the MINERVA-Red ECS. Diagram is to scale with the exception of the size of the RTDs}
\label{fig:diagram}
\end{figure}

Heat is generated inside the Thermal Enclosure using two heater units that consist of two 15$\Omega$ thin-film resistors connected in series and are directly mounted on a CPU heat sink and fixed in front of a 10" Caravek DC CD24B3 fan run at 15 volts. The Caravek fan is capable of moving 550 cubic feet per minute at 24 volts. We operate the fans at a lower voltage because our Thermal Enclosure has a small volume of only five square feet and the fans also generate more heat then we want when operated at 24 volts. Each heater unit is driven by a DC supply that provides up to 50 watts of heat. Together, the heater units are able to maintain the Thermal Enclosure more than 5$^{\circ}$C above the room ambient. We note that the fans themselves, which are necessary to homogenize the temperature within the Thermal Enclosure, generate enough heat to maintain the Thermal Enclosure at 4$^{\circ}$C above room ambient. An  example of a heater unit is shown in Figure \ref{fig:heater}, and the locations of the heater units within the Thermal Enclosure are shown in Figure \ref{fig:diagram}. Since the CCD camera itself is a heat source, which we discuss further in Section 4.1, we positioned one heater unit in close proximity to the CCD, and the other on the opposite side of the Thermal Enclosure.

The control unit for the heaters is a Stanford Research Systems (SRS) PTC10 with  one four-channel, four-wire platinum Resistance Temperature Device (RTD) reader card and two 50W DC output channels to drive the two heater units inside the Thermal Enclosure. The accuracy of the PTC10 for a 100 $\Omega$ RTD is $\pm{0.008}\Omega$, the drift is $\pm{0.001}\Omega$/$^{\circ}$C, and the noise (RMS) is 0.0003$\Omega$\footnote{http://www.thinksrs.com/downloads/PDFs/Catalog/PTC10c.pdf}. Four-wire RTDs produced by Omega were chosen to read the air temperatures inside the Thermal Enclosure (Omega part number P-L-1/3-1/8-3-0-P-3). These RTDs are specifically designed  for high-precision and high-accuracy and are packaged for measuring air temperature. The PTC10 has a sophisticated autotuning capability that dramatically simplifies the job of setting the PID control loop parameters. To achieve the highest precision possible the total resistance of the heater units should be near 30$\Omega$. If the resistance is too low or too high, the PTC10 DC cards can overheat causing the system to turn off and put the unit in a failsafe mode.

In order to maximize the efficiency of the heater units and PID control, and to accommodate any large variations in room ambient temperature, we set the PTC10 setpoint to 5$^{\circ}$C above the average Thermal Enclosure temperature when all heat sources are on. This includes the fans and camera with the CCD cold. Typically, our laboratory environment is 20.0$^{\circ}$C $\pm0.75^{\circ}$C, and turning the heater unit fans on increases the Thermal Enclosure temperature by 4$^{\circ}$C. Based on this, we chose to operate the ECS at a setpoint of 30.0$^{\circ}$C, approximately 10$^{\circ}$C above room ambient. We note that if the room temperature were to increase to 26$^{\circ}$C, the ECS would fail as it has no active cooling capability.

Additional temperature monitoring channels are provided by OMEGA PT-104a temperature loggers connected to 4-wire 100 $\Omega$ RTDs. The PT-104a logger has an advertised precision of 1~mK and an accuracy of 10~mK for 100$\Omega$ or 1000 $\Omega$ RTDs\footnote{http://www.omega.com/das/pdf/PT-104A.pdf}. Together we have twelve logger channels placed both inside the Thermal Enclosure and inside the vacuum chamber to provide independent temperature data from the PTC10. Each channel is recorded once per second.
We tested the stability of the temperature-measuring electronics by attaching a 100$\Omega$ ultrastable resistor to the Omega PT-104a. A resistor of exactly 100$\Omega$ should read zero $^{\circ}$C. The internal resistance changes 0.05 ppm/$^{\circ}$C or 6 $\mu$K per~$^{\circ}$C, far below our measurement ability. Over a week-long test, 95.5\% of the recorded temperature measurements were within $\pm{2}$~mK of zero degrees C. We can therefor conclude that variations larger than 2~mK are real and can be measured effectively using our apparatus. Anything under a 2~mK change could be attributed to electronic noise from the PT-104a. It should also be noted that the PT100 can drift with time in addition to exhibiting a hysteresis effect \citep{PTstability}. However, Omega Engineering has no related data for any of the PT100s they produce. Acquiring such data for the sensors used here was outside of the scope of this work.  

%\begin{deluxetable}{ccccccccc}
%\tabletypesize{\scriptsize}

\begin{deluxetable}{ c c c c }[h]
\tablecaption{Budget for the ECS}
 \startdata
 \\
 Item & Cost per Item & Quanitity & Total \\
 Standford Research Sytems PTC 10 & \$1500 & 1 & \$1500 \\
 Standford Research Systems 4-channel Pt RTD card & 600 & 1 & \$600 \\ 
 Standford Research Systems 50 W DC output card & \$500 & 2 & \$1000  \\  
 Omega PT-104a & \$700 & 3 & \$2100  \\ 
 Solid State Cooling Oasis 3 Liquid Chiller (three decimal readout) & \$2900 & 1 & \$2900 \\
 Caravek DC CD24B3 Fan & \$90 & 2 & \$180\\
 Omega P-L-1/3-1/8-3-O-P-3 RTD & \$80 & 6 &\$420 \\
 Omega PR-20-3-100-1/4-2-E-T RTD & \$80 & 3 & \$240 \\
 Omega F2222-100-1/3B-100 RTD & \$3.65 & 100  &\$365 \\
 Masterkleer PVC Tubing (Mcmaster part number 5233K56) & \$0.24 per foot & 50' & \$12 \\
 Koolance 702 Liquid Coolant 750 ml & \$15 & 3 & \$45 \\
 Ridged Foam insulation 4'x8'x2" R=10 & \$30 & 4 & \$120 \\
 Slit Foam Rubber Pipe Insulation (Mcmaster part number 44745K47) & \$13 & 4 & \$52 \\
 Startech 95mm Socket T 775 CPU Cooler Fan with Heatsink & \$15 & 2 &\$30
 \\
 Approximate Total & & & \$9600
\\
\enddata
\end{deluxetable}

\newpage
\subsection{Performance Limitations}

Even if the heater units and PID temperature controller were to do a perfect job controlling the temperature of the air inside the Thermal Enclosure, there exist conductive and radiative paths that could couple the optical components inside the vacuum chamber to variations in the outside temperature. In our ECS, there are three primary conductive paths that we consider and have taken steps to minimize.

The first is the CCD detector, which is thermoelectrically cooled. MINERVA-Red uses an Andor Ikon-L 936 BR-DD detector, which is directly attached to the vacuum chamber and is inside the Thermal Enclosure. To reduce dark current in the CCD, the device is operated at $-85.0^{\circ}$C, which is achieved using a five-stage Peltier cooler. This cooling processes necessarily generates heat (many watts, in this case), which is deposited inside the Thermal Enclosure. To mitigate this effect, we use liquid coolant to move the majority of the heat generated by cooling the CCD to outside the Thermal Enclosure. Since the efficiency of the CCD cooler, and therefore the amount of heat it generates, depends on the liquid coolant temperature, our application requires a chiller capable of maintaining extremely precise control of the temperature of the coolant.

\begin{figure}[htp]
\centering
\includegraphics[width=7cm, angle = 270, trim={4cm 0cm 4cm 0cm},clip]{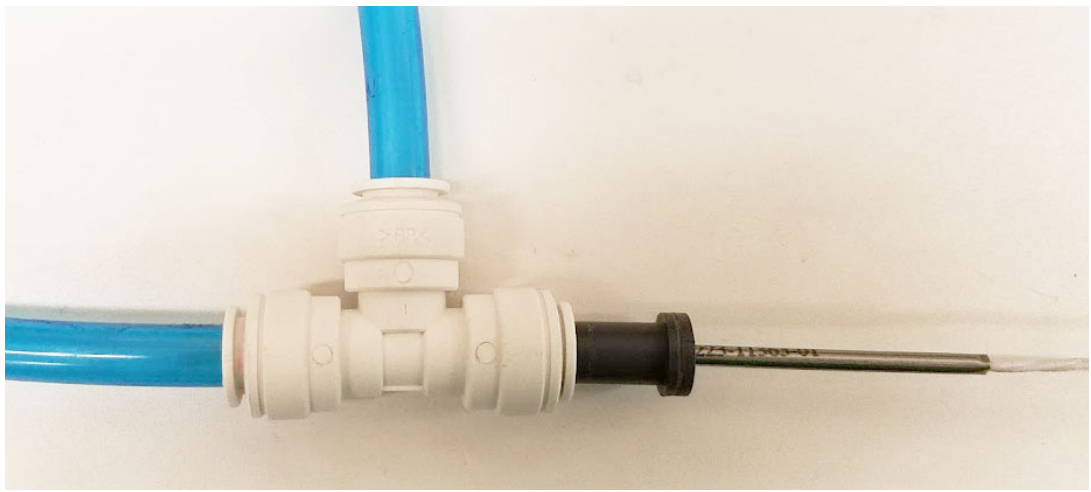}
\caption{An Omega PR-20-2-100-1/4-2-E-T RTD is installed inside of a plastic bushing, which is inserted into a plastic T to control and monitor the liquid coolant. This packaging allows for the RTD to be submersed in liquid. The sensor is 2" long and is friction fit inside the plastic bushing. The first inch in inserted into the liquid. However, additional units were purchased and placed downstream of the chiller to control and monitor the liquid temperature. A three-wire RTD is used for control with the Oasis 3 as that is the only type of input available. We note that in the picture the RTD is not inserted into the bushing all the way, nor is the bushing inserted into the T all the way.}
\label{fig:liquirdrtd}
\end{figure}

The Oasis 3 chiller manufactured by Solid State Cooling can maintain liquid temperature stability of $\pm 10$~mK, the best performance of a commercially available water chiller. We modified the Oasis 3 operation slightly by placing the control temperature sensor, a PT100 RTD, downstream in the liquid supply line to the CCD, just inside the Thermal Enclosure. Figure \ref{fig:liquirdrtd} shows how the sensor is inserted into the liquids path. Without this modification, the Oasis 3 would be controlling the temperature of the coolant leaving the unit, which would then be subject to temperature modulations as it traverses the 2 meter distance between the chiller and the Thermal Enclosure. For the MINERVA-Red ECS, we found that the Oasis 3 with downstream temperature feedback provides liquid temperature stability of $\pm 5$~mK over weeks. We use Coolance liq-702 as a coolant which features a mixture of glycol (25$\%$) and water (75$\%$). We insulate the $\frac{1}{4}"$ PVC tubing running from the chiller to the CCD and back using foam pipe insulation, taking care to seal the ends of the pipe insulation, where the PVC tubing enters or exits the insulation, with polyurethane spray foam to provide a sealed path for the liquid between the chiller and the CCD. Similar insulating steps were taken where the coolant tubing enters and exits the Thermal Enclosure. T shaped pieces of insulation were employed over the temperature sensors to ensure they measured the liquid temperature and not the air temperature.

In addition to the impact of water temperature on the heat output of the Peltier coolers, we also found that the temperature of liquid coolant directly impacts the temperature of the CCD camera housing, which is in direct contact with the vacuum chamber. Changing the coolant setpoint with the CCD cooler off produces a rapid change, of similar amplitude, in the temperature of the vacuum chamber wall adjacent to the CCD camera. To mitigate this effect, we machined a custom G10 spacer between the CCD face plate and the mounting flange on the vacuum chamber. G10 is not strong enough to cut a knife edge in copper to create the classic vacuum seal. We therefore replaced the copper gaskets with Viton o-rings. Since our vacuum level is on the order of 1 mTor, Viton is a viable alternative to copper.

A second conductive path stems from the fact that the vacuum chamber is physically mounted to the optical table below. Three PEEK standoffs $\frac{3}{4}"$ in diameter and 4" long connect the chamber to the optical table. There are three mounting brackets on the chamber that were designed for this purposes. The PEEK standoffs are tapped on both ends. The bottom of the standoffs are threaded into the optical table using set screws. The chamber rests on the top of the cylinders and is connected by a socket head cap screw. The length of the posts were set so that a  1" air gap exists between the Thermal Enclosure and the optical table, as well as a 1" air gap between the bottom of the vacuum chamber and the foam board of the Thermal Enclosure. The holes that are drilled in the Thermal Enclosure to accommodate the mounting posts are back filled with polyurethane spray foam. The PEEK material was chosen for low its thermal conductivity and because it is relatively inexpensive and can be machined using standard tools.

To evaluate the magnitude of the possible differential heat load between the vacuum chamber inside the Thermal Enclosure and the optical bench below we can employ Fourier's Law of heat conduction,
\begin{equation}
\frac{Q}{t} =\frac{\kappa A(T_{\mathrm{Hot}} - T_{\mathrm{Cold}})}{L} 
\end{equation}
where $\frac{Q}{t}$ is the rate of heat added or removed from the system (in units of Watts), $\kappa$ is the thermal conductivity of PEEK (0.25W m$^{-1}$ K$^{-1}$), $A$ is the cross sectional area of the three PEEK feet (0.00085 m$^{2}$), and $L$ is the length of the PEEK feet (0.102 m). For a $\Delta$T of 10~$^{\circ}$C the heat flow rate, for three feet, is 0.02088~W from the vacuum chamber to the cooler laboratory environment.

We assume that the vacuum chamber is in thermal equilibrium, so that the flow of heat out of the vacuum chamber to the outside environment is balanced by heat flow into the vacuum chamber from the Thermal Enclosure, which is actively heated. We approximate the bottom plate of the vacuum chamber as having an area of 0.456 m$^{2}$, a thickness of 1.3 cm, and a thermal conductivity appropriate for Al 6061(200 W m$^{-1}$~K$^{-1}$). The density of AL 6061 is 2700~Kg~m$^{-3}$ which means the weight of the base plate of the chamber is 23.39 Kg. The specific heat of Al 6061 is 896 J~Kg$^{-1}$~K$^{-1}$. To calculate the heat capacity of the plate we simply multiply the weight by the specific heat to find Q = 20957~J~K$^{-1}$.

To determine the change in the temperature of the bottom of the vacuum chamber we can divide the heat flow rate through the PEEK feet by the heat capacity of the Al base plate of the vacuum chamber. This is approximately 100 $\mu$K per second with $\Delta T=10$~K across the three PEEK feet. This means that after six hours of a constant laboratory temperature offset, the expected change in the temperature of the base of the optical chamber would be just large enough for us to measure ($\Delta T$=2~mK). There are other conductive paths between the Minerva-Red vacuum chamber and the laboratory environment, such as the vacuum hose and electrical wires. However, we expect the contributions from these paths to be small since thermal variations are damped by having a length of hose or wire inside the temperature stabilized environment of the Thermal Enclosure. Based on our calculations, we conclude that that the CCD camera and its coolant supply are likely the primary thermally conductive path impacting the overall performance of the ECS.

\newpage
\subsection{Passive Radiation Shield}
\begin{figure}[htp]
\centering
\includegraphics[width=8cm]{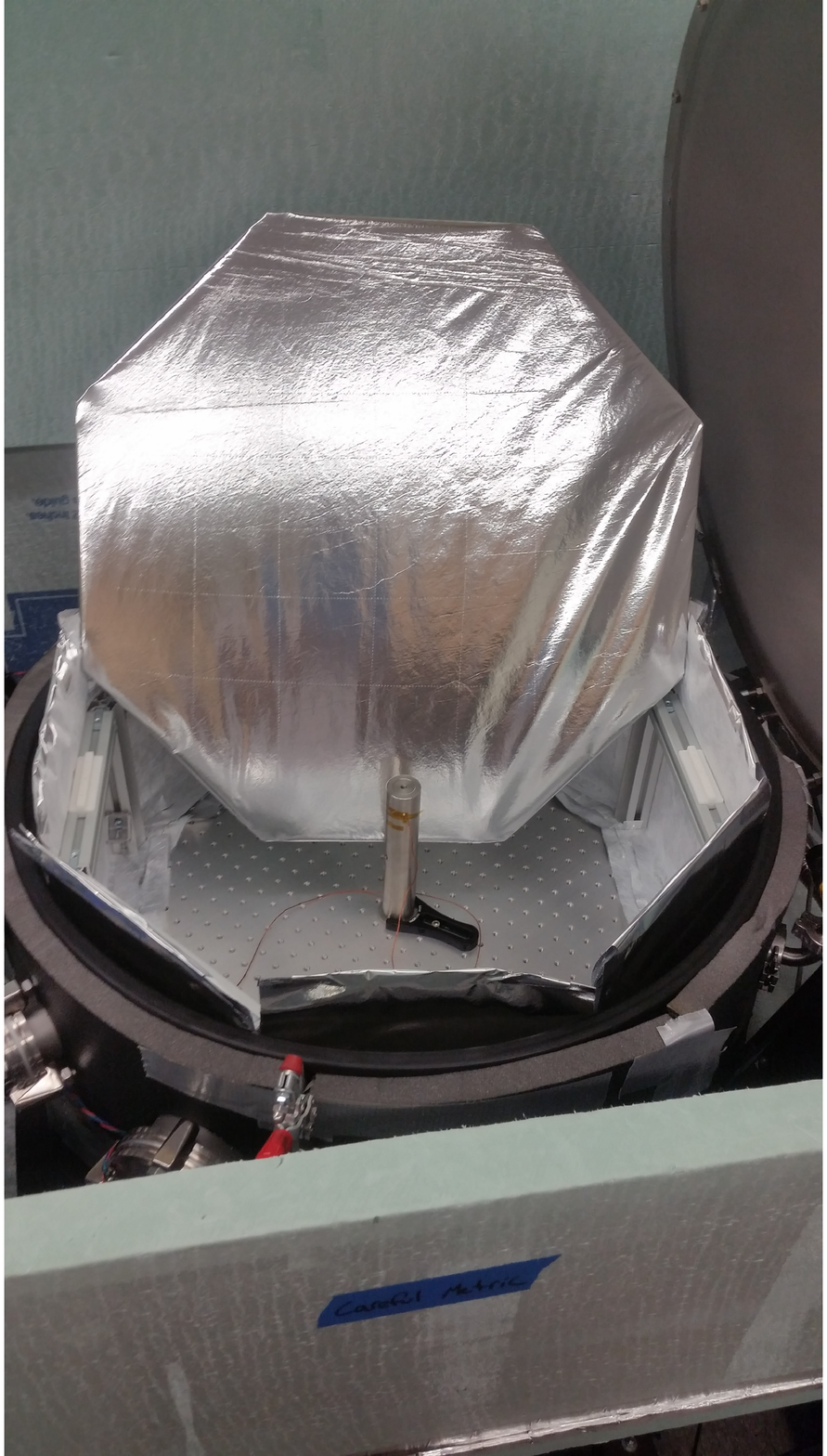}
\includegraphics[width=8cm]{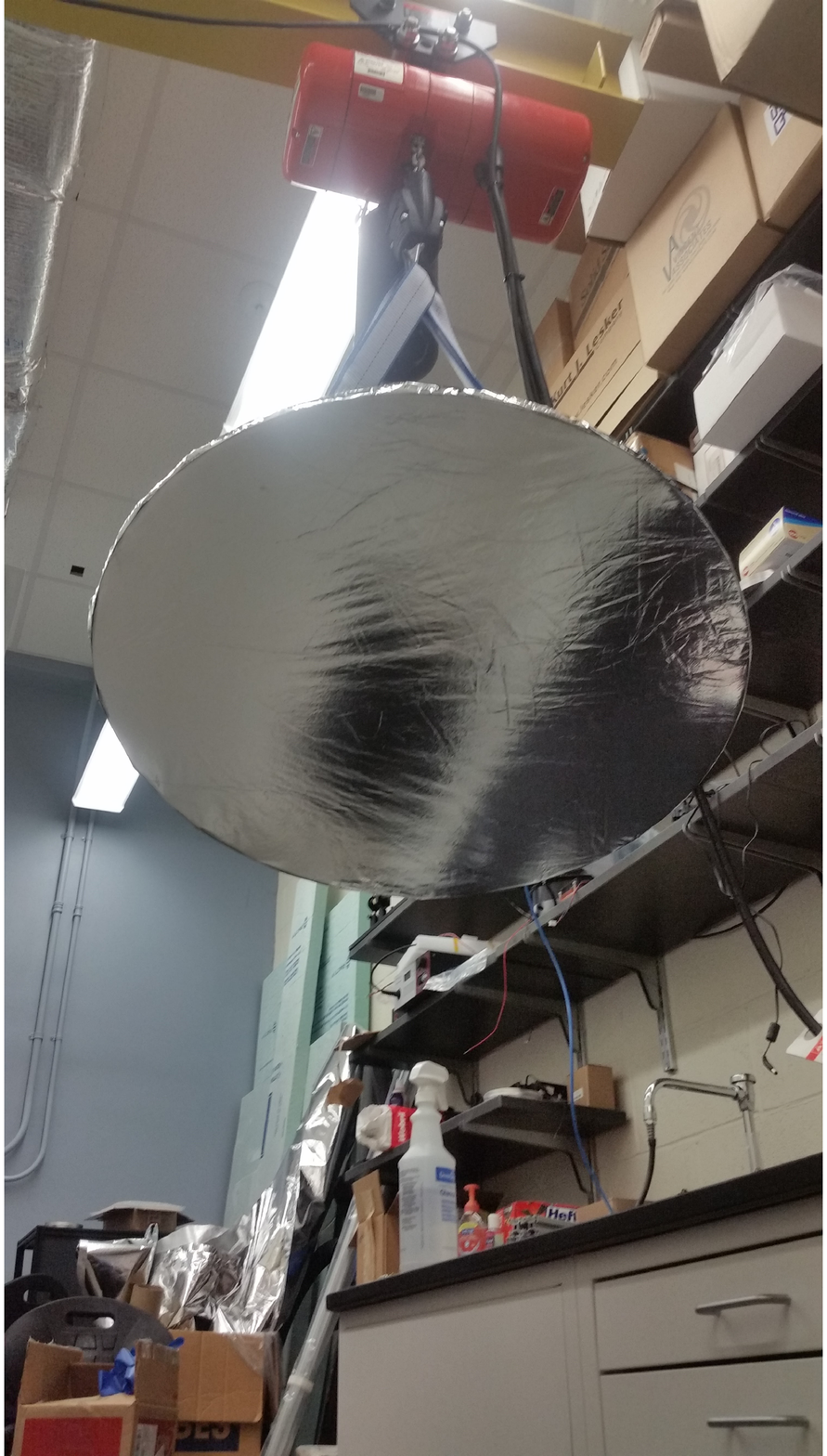}
\caption{The MINERVA-Red MLI tent being installed inside the vacuum chamber. The layers of mylar in the MLI material provide an effective barrier to radiative coupling between the vacuum chamber walls and the optical board inside the chamber.}
\label{fig:mli}
\end{figure}

As a result of the thermal shorts discussed above, we found that the thermal stability inside the vacuum chamber was worse than the stability of the air inside the Thermal Enclosure, with variations in the temperatures of the vacuum chamber walls several~mK larger than variations in the air temperature inside the Thermal Enclosure. With the vacuum chamber operating at a pressure of 3.5 mTorr, convective thermal coupling between the vacuum chamber walls and the optical components inside the chamber is greatly reduced. However, there is a radiative coupling that we mitigated using a tent of Multi-Layer Insulation (MLI) as a radiation shield to decouple the interior of the vacuum chamber from variations in the temperatures of the chamber walls. An extruded aluminum frame was designed and wrapped with six layers of MLI and placed over, around, and beneath the optical bench inside the vacuum chamber, as shown in Figure \ref{fig:mli}. The MLI was manufactured by MEI-Thermolam for the Balloon-borne Large Aperture Submillimeter Telescope (BLAST)  \citep{BLAST}. The MLI is 150$\mu$m thick mylar with 40 nm thick aluminum coatings. It has a 1.2$\mu$m thick reemay (0.5 oz per square yard) spunbonded on the back of the mylar. We estimate this reduces the amount of radiative coupling between the chamber walls and optical elements by a factor of four. This provides an optically thick and reflective surface at optical to millimeter wavelengths.

\section{Results}
We tested the MINERVA-Red ECS under nominal laboratory conditions over a period of three months. Here, we report the results of a five-week test where the laboratory room temperature was allowed to oscillated under normal HVAC conditions. People entered the room over this period of time, which caused the HVAC system to react, as did large swings in in day-to-day weather. Two unplanned events happened during our test. On day 17 the room decreased 1.0$^{\circ}$C. On day 22 cooling to the HVAC system was inadvertently disabled. This resulted in a 1.5$^{\circ}$C increase in the laboratory temperature. These two events will serve as stress tests for the performance of the ECS.

\begin{figure}[h]
\centering
\includegraphics[width=12cm, trim={0cm 6cm 0cm 6cm},clip]{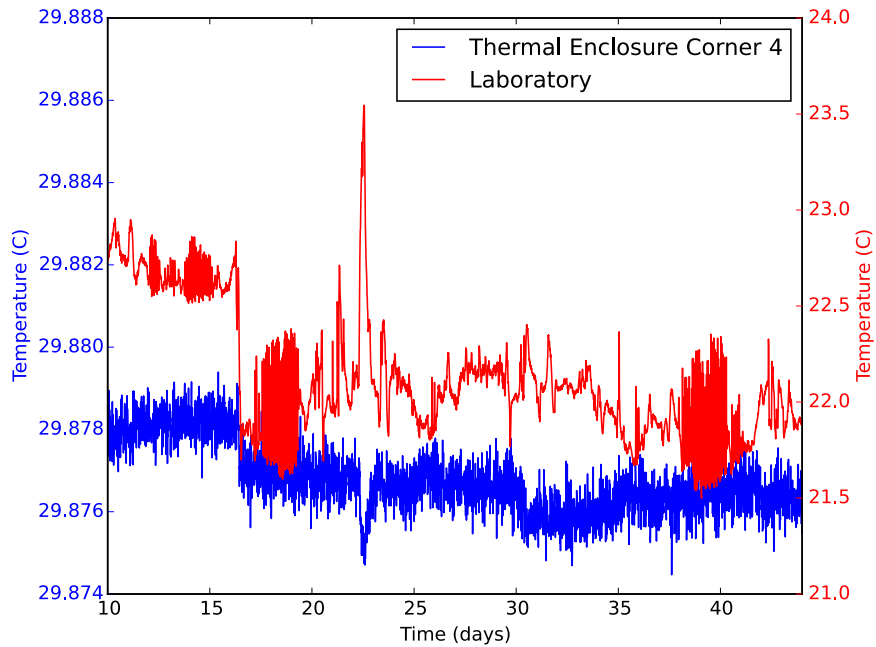}
\caption{Temperature of the laboratory environment compared to the ambient temperature inside the Thermal Enclosure with 15 minute binning. Around day 22 there was a failure of the HVAC system in the laboratory, leading to a sharp increase in the room temperature. The resulting impact on the Thermal Enclosure temperature was less than 1~mK. Note that the scales on the two Y axes are different. }
\label{fig:roomvbox}
\end{figure}

\begin{figure}[h]
\centering
\includegraphics[width=12cm,trim={0cm 6cm 0cm 6cm},clip]{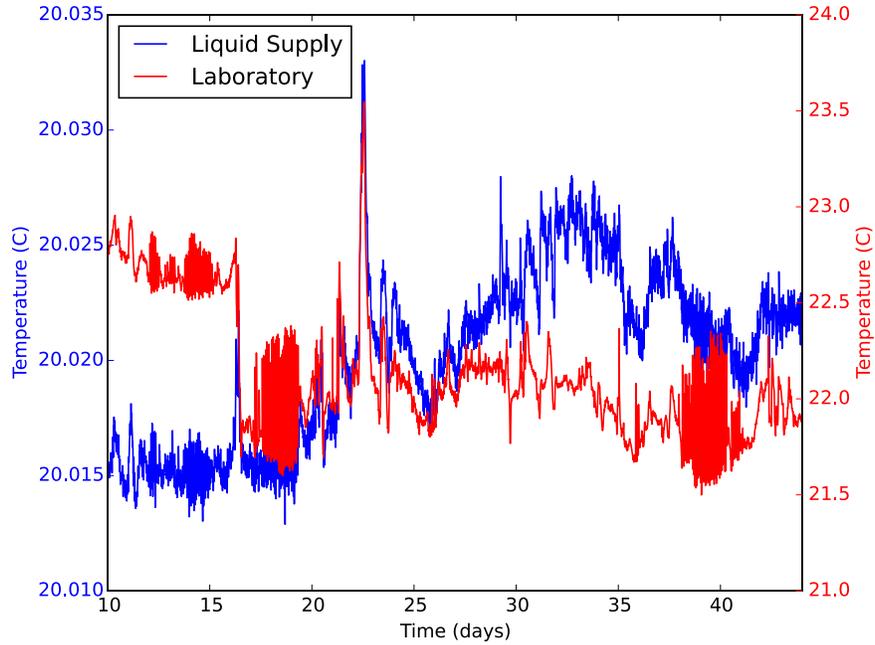}
\caption{Temperature of the CCD coolant entering the Thermal Enclosure compared to the laboratory temperature with 15 minute binning. Note that the scales on the two Y axes are different.}
\label{fig:liquidsupply}
\end{figure}

Temperatures of the room compared to the air temperature inside Thermal Enclosure are shown in Figure \ref{fig:roomvbox}. Over 35 days, the room temperature varies by up to 2.0$^{\circ}$C (PV) when averaged over 15-minute bins. Over this same period and binning, the Thermal Enclosure environment varies by 6~mK (PV), with variations less than $\pm1.0$~mK seen in over 88 percent of the 15-minute bins. In Figure \ref{fig:liquidsupply} we can see that the liquid coolant for the CCD is not completely isolated from changes in room temperature.  We can see a 20~mK drift in temperature of the coolant supply over the 35-day period. As expected, the return coolant temperature is on average 0.5$^{\circ}$C warmer than the supply, since the coolant is absorbing heat from the CCD cooler.

\begin{figure}[h]
\centering
\includegraphics[width=12cm,trim={0cm 6cm 0cm 6cm},clip]{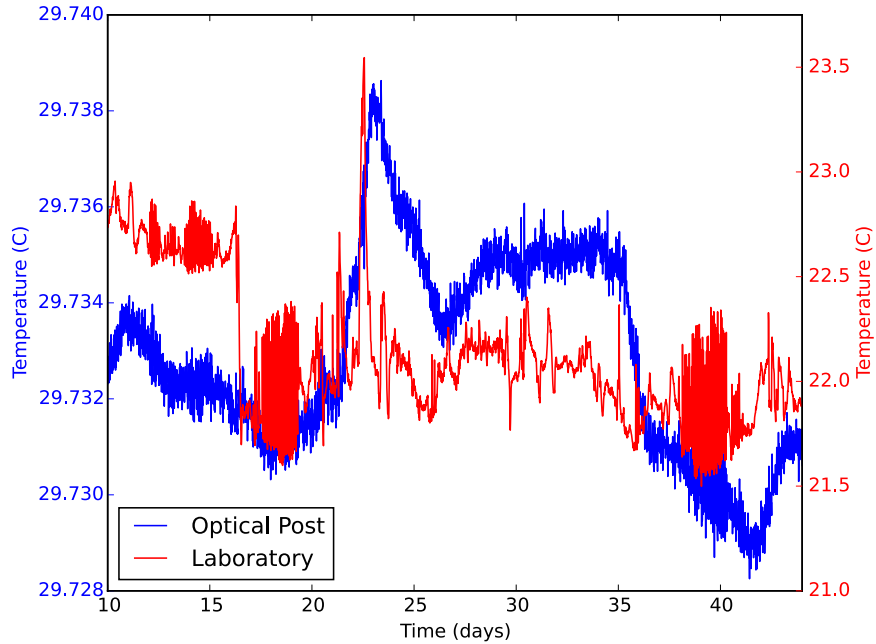}
\caption{Temperature of the laboratory compared to the temperature of an optical post mounted on the optical board within the vacuum chamber when pumped down. The binning of the data is 15 minute binning. The large increase in laboratory temperature due to an HVAC failure around day 22 resulted in a 6~mK increase in the temperature of the optical post lasting for nearly five days. Note that the scales on the two Y axes are different.}
\label{fig:roompost}
\end{figure}

Using RTDs placed inside the vacuum chamber, we measure the temperature variations that optical components such as the Echelle grating would have experienced over this 35-day test. As a proxy for grating temperature, we monitor an RTD attached to a 1.5" diameter optical post mounted in the center of the evacuated vacuum chamber. Once the Thermal Enclosure is sealed and the temperature control system is turned on, it takes approximately 10 days for the temperature of the optical post to equilibrate with the temperature inside the Thermal Enclosure. This is why Figures \ref{fig:roomvbox}, \ref{fig:liquidsupply}, and \ref{fig:roompost} start at day 10. In Figure \ref{fig:roompost} we see that the temperature of the optical post varies by $\pm{5}$~mK (PV), surpassing our initial design goal of $\pm{10}$~mK.

On day 17, the room decreased 1.0$^{\circ}$C in mean temperature. We observe a six hour increase in liquid coolant temperature of 5~mK. Once the room stabilizes the liquid coolant stabilizes at its previous temperature. We also observe a 1~mK decrease in the Thermal Enclosure temperature over the same period. This results in a 1~mK decrease in the optical post temperature. We can conclude that a 1.0$^{\circ}$C decrease in laboratory temperature has a negligible effect on optical components inside the vacuum chamber.

On day 22, the room increased 1.5$^{\circ}$C for six hours due to an HVAC failure. We observe a 12~mK increase in the liquid supply lasting six hours. The response of the optical post to this temperature increase is shown in Figure \ref{fig:roompost}. The post increases in temperature 5~mK over 19 hours then slowly decreases in temperature over 60 hours. We emphasize that the laboratory temperature increase resulting from the HVAC failure is larger and more rapid than the temperature fluctuations that the MINERVA-Red spectrometer will experience in the spectrograph room at Mt. Hopkins, yet the ECS is still able to maintain instrumental temperature stability better than $\pm10$~mK. The results of the temperature stress tests indicate that the thermal coupling between the optical components inside the MINERVA-Red vacuum chamber, which is inside the Thermal Enclosure, are more complicated than anticipated. For example, the calculations presented in Section 4.1 indicate that the response of the vacuum to just the change in the temperature of the underlying optical table (which is at laboratory temperature) should have been much smaller than the 6~mK we observed. Based on our experience with the MINERVA-Red ECS, we believe that it is the control of the CCD liquid supply temperature that is currently limiting our ability to isolate optical components inside the vacuum chamber from the outside environment. Independent of large variations in the laboratory temperature, we observe few-mK variations in the liquid temperature that correlate with changes in the optical post temperature. At this time, we do not believe that there is a better commercially available solution for controlling the temperature of the liquid used to cool the MINERVA-Red CCD.

\begin{figure}[h]
\centering
\includegraphics[width=12cm,trim={0cm 6cm 0cm 6cm},clip]{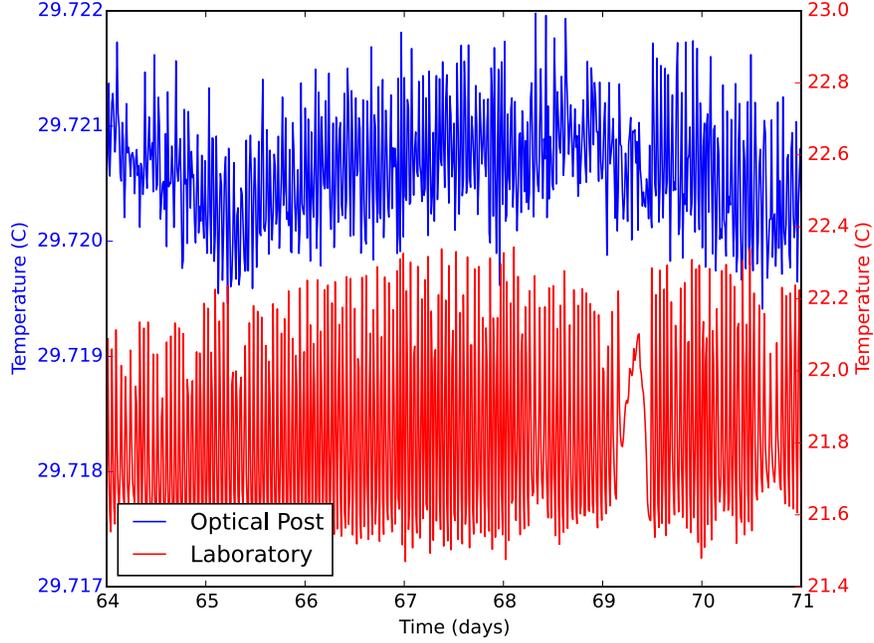}
\caption{Temperature of the laboratory compared to the temperature of the optical post mounted to the optical board within the vacuum chamber when pumped down. The binning of the data is 15 minutes. This plot shows the room stable to $\pm{0.45}$C which allows the post to be stable to $\pm{1.5}$~mK.}
\label{fig:roompoststable}
\end{figure}

After 35 days of tests with the CCD camera cold but not integrating or reading out, we simulated spectroscopic data collection by taking 15-minute dark frames constantly for seven consecutive days. Initially, we saw a 3~mK decrease in the camera body temperature. According to the camera manufacturer, this is due to the fact that the CCD electronics actually consume less power when the CCD is integrating compared to when it is idle or reading out. Once the vacuum chamber equalized to this new camera body temperature, we observed no noticeable difference in how the optical post reacted to small temperature perturbations in the room or liquid coolant temperatures while the CCD was collecting dark frames. Since we know the camera has two modes of heat dissipation, with MINERVA-Red we plan to constantly take exposures, either science frames or dark frames, both at night and during the day.

Starting approximately 64 days after the start of these tests, the laboratory was left untouched for a week. This resulted in a room stability of 0.9$^{\circ}$C PV over the course of the seven days. With the laboratory stable, the optical post temperature was stable to 2.6$^{\circ}$~mK PV as seen in Figure  \ref{fig:roompoststable}. This suggests that in operating conditions similar to what we might expect at Mt. Hopkins, or in any well controlled and isolated room, we can expect $\pm{1.5}$~mK stability over long time periods. 

\section{Conclusion}
Using inexpensive, ``off-the-shelf" components we have built and demonstrated an Environmental Control System (ECS) for the MINERVA-Red spectrograph. By using resistive heaters and a PID controller to maintain the instrument temperature above room ambient, we achieve a maximum temperature variation of $\Delta T=10$~mK within our vacuum chamber while the laboratory environment varies by up to 1.5$^{\circ}$C. When the laboratory environment is stable with temperature oscillations less than 1.0$^{\circ}$C, we find that the temperature stability inside the chamber is $\Delta T=3$~mK. This level of temperature stability is comparable to that demonstrated by existing Doppler spectrometers capable of 1 m s$^{-1}$ radial velocity measurement precision. We investigate strategies for handling the heat dissipated by a thermoelectrically cooled CCD camera and find that liquid coolant with a commercial water chiller is sufficient for minimizing variations in the heat output of the CCD. However, we found that the heat output from the CCD electronics is different in idle or integrating modes. Based on a simple estimate of the effect of temperature variations on an Echelle grating, we show that this level of temperature stability is a minimum requirement to enable intrinsic instrumental stability better than 1 m s$^{-1}$. While this ECS is specifically designed for the MINERVA-Red spectrometer, the approach described here should be applicable to a wide range of small astronomical instruments requiring a thermally stable operating environment, but not cryogenic temperatures.

\section{acknowledgements} We would like to thank an anonymous referee whose comments helped to substantially improve this manuscript. We would like to thank Jeff Klein of the University of Pennsylvania for his many helpful comments on this manuscript and for his assistance with the construction and design of the passive radiation shield. This work was supported by NASA through a Nancy Grace Roman Technology Fellowship to CHB. This work was performed, in part, by SPH under contract with the Jet Propulsion Laboratory (JPL) funded by NASA through the Sagan Fellowship Program executed by the NASA Exoplanet Science Institute.

%%%%%%%%%%%%%%%%%%%%%%%%%%%
%REFERENCES

%\bibliography{ECSbib.bib}

\begin{thebibliography}{}
\expandafter\ifx\csname natexlab\endcsname\relax\def\natexlab#1{#1}\fi

\bibitem[{{Blake} {et~al.}(2015){Blake}, {Johnson}, {Plavchan}, {Sliski},
  {Wittenmyer}, {Eastman}, \& {Barnes}}]{MINERVARed}
{Blake}, C., {Johnson}, J., {Plavchan}, P., {et~al.} 2015, in American
  Astronomical Society Meeting Abstracts, Vol. 225, American Astronomical
  Society Meeting Abstracts, 257.32

\bibitem[{{Crepp} {et~al.}(2016){Crepp}, {Crass}, {King}, {Bechter}, {Bechter},
  {Ketterer}, {Reynolds}, {Hinz}, {Kopon}, {Cavalieri}, {Fantano}, {Koca},
  {Onuma}, {Stapelfeldt}, {Thomes}, {Wall}, {Macenka}, {McGuire}, {Korniski},
  {Zugby}, {Eisner}, {Gaudi}, {Hearty}, {Kratter}, {Kuchner}, {Micela},
  {Nelson}, {Pagano}, {Quirrenbach}, {Schwab}, {Skrutskie}, {Sozzetti},
  {Woodward}, \& {Zhao}}]{iLocator}
{Crepp}, J.~R., {Crass}, J., {King}, D., {et~al.} 2016, in \procspie, Vol.
  9908, Ground-based and Airborne Instrumentation for Astronomy VI, 990819

\bibitem[{{Feger} {et~al.}(2014){Feger}, {Bacigalupo}, {Bedding}, {Bento},
  {Coutts}, {Ireland}, {Parker}, {Rizzuto}, \& {Spaleniak}}]{RHEA}
{Feger}, T., {Bacigalupo}, C., {Bedding}, T.~R., {et~al.} 2014, in \procspie,
  Vol. 9147, Ground-based and Airborne Instrumentation for Astronomy V, 91477I

\bibitem[{{Galitzki} {et~al.}(2014){Galitzki}, {Ade}, {Angil{\`e}}, {Ashton},
  {Beall}, {Becker}, {Bradford}, {Che}, {Cho}, {Devlin}, {Dober}, {Fissel},
  {Fukui}, {Gao}, {Groppi}, {Hillbrand}, {Hilton}, {Hubmayr}, {Irwin}, {Klein},
  {van Lanen}, {Li}, {Li}, {Lourie}, {Mani}, {Martin}, {Mauskopf}, {Nakamura},
  {Novak}, {Pappas}, {Pascale}, {Pisano}, {Santos}, {Savini}, {Scott},
  {Stanchfield}, {Tucker}, {Ullom}, {Underhill}, {Vissers}, \&
  {Ward-Thompson}}]{BLAST}
{Galitzki}, N., {Ade}, P.~A.~R., {Angil{\`e}}, F.~E., {et~al.} 2014, Journal of
  Astronomical Instrumentation, 3, 1440001

\bibitem[{{Ge} {et~al.}(2016){Ge}, {Ma}, {Muterspaugh}, {Singer}, {Varosi},
  {Powell}, {Williamson}, {Sithajan}, {Grieves}, {Zhao}, {Schofield}, {Liu},
  {Cassette}, {Carlson}, {Klanot}, {Jeram}, \& {Barnes}}]{Dharma}
{Ge}, J., {Ma}, B., {Muterspaugh}, M.~W., {et~al.} 2016, in American
  Astronomical Society Meeting Abstracts, Vol. 227, American Astronomical
  Society Meeting Abstracts, 220.06

\bibitem[{{Hall} {et~al.}(2016){Hall}, {Plavchan}, {Geneser}, {Giddens}, \&
  {Spangler}}]{MICRONERVA}
{Hall}, R., {Plavchan}, P., {Geneser}, C., {Giddens}, F., \& {Spangler}, S.
  2016, in American Astronomical Society Meeting Abstracts, Vol. 228, American
  Astronomical Society Meeting Abstracts, 316.09

\bibitem[{{Halverson} {et~al.}(2014){Halverson}, {Mahadevan}, {Ramsey},
  {Hearty}, {Wilson}, {Holtzman}, {Redman}, {Nave}, {Nidever}, {Nelson},
  {Venditti}, {Bizyaev}, \& {Fleming}}]{SAMFIBER}
{Halverson}, S., {Mahadevan}, S., {Ramsey}, L., {et~al.} 2014, \pasp, 126, 445

\bibitem[{{Halverson} {et~al.}(2016){Halverson}, {Terrien}, {Mahadevan}, {Roy},
  {Bender}, {Stef{\'a}nsson}, {Monson}, {Levi}, {Hearty}, {Blake}, {McElwain},
  {Schwab}, {Ramsey}, {Wright}, {Wang}, {Gong}, \& {Roberston}}]{SAMNEID}
{Halverson}, S., {Terrien}, R., {Mahadevan}, S., {et~al.} 2016, in \procspie,
  Vol. 9908, Ground-based and Airborne Instrumentation for Astronomy VI, 99086P

\bibitem[{{Jovanovic} {et~al.}(2016){Jovanovic}, {Schwab}, {Cvetojevic},
  {Guyon}, \& {Martinache}}]{Subaru}
{Jovanovic}, N., {Schwab}, C., {Cvetojevic}, N., {Guyon}, O., \& {Martinache},
  F. 2016, \pasp, 128, 121001

\bibitem[{{Jovanovic} {et~al.}(2017){Jovanovic}, {Schwab}, {Guyon}, {Lozi},
  {Cvetojevic}, {Martinache}, {Leon-Saval}, {Norris}, {Gross}, {Doughty},
  {Currie}, \& {Takato}}]{NEM}
{Jovanovic}, N., {Schwab}, C., {Guyon}, O., {et~al.} 2017, ArXiv e-prints,
  arXiv:1706.08821

\bibitem[{{Ljungblad} {et~al.}(2013){Ljungblad}, {Holmsten}, {Josefson}, \&
  {Klevedal}}]{PTstability}
{Ljungblad}, S., {Holmsten}, M., {Josefson}, L.-E., \& {Klevedal}, B. 2013, in
  American Institute of Physics Conference Series, Vol. 1552, American
  Institute of Physics Conference Series, ed. C.~W. {Meyer}, 421--426

\bibitem[{{Mayor} {et~al.}(2003){Mayor}, {Pepe}, {Queloz}, {Bouchy},
  {Rupprecht}, {Lo Curto}, {Avila}, {Benz}, {Bertaux}, {Bonfils}, {Dall},
  {Dekker}, {Delabre}, {Eckert}, {Fleury}, {Gilliotte}, {Gojak}, {Guzman},
  {Kohler}, {Lizon}, {Longinotti}, {Lovis}, {Megevand}, {Pasquini}, {Reyes},
  {Sivan}, {Sosnowska}, {Soto}, {Udry}, {van Kesteren}, {Weber}, \&
  {Weilenmann}}]{HARPS}
{Mayor}, M., {Pepe}, F., {Queloz}, D., {et~al.} 2003, The Messenger, 114, 20

\bibitem[{{Pepe} {et~al.}(2010){Pepe}, {Cristiani}, {Rebolo Lopez}, {Santos},
  {Amorim}, {Avila}, {Benz}, {Bonifacio}, {Cabral}, {Carvas}, {Cirami},
  {Coelho}, {Comari}, {Coretti}, {De Caprio}, {Dekker}, {Delabre}, {Di
  Marcantonio}, {D'Odorico}, {Fleury}, {Garc{\'{\i}}a}, {Herreros Linares},
  {Hughes}, {Iwert}, {Lima}, {Lizon}, {Lo Curto}, {Lovis}, {Manescau},
  {Martins}, {M{\'e}gevand}, {Moitinho}, {Molaro}, {Monteiro}, {Monteiro},
  {Pasquini}, {Mordasini}, {Queloz}, {Rasilla}, {Rebord{\~a}o}, {Santana
  Tschudi}, {Santin}, {Sosnowska}, {Span{\`o}}, {Tenegi}, {Udry}, {Vanzella},
  {Viel}, {Zapatero Osorio}, \& {Zerbi}}]{ESPRESSO}
{Pepe}, F.~A., {Cristiani}, S., {Rebolo Lopez}, R., {et~al.} 2010, in
  \procspie, Vol. 7735, Ground-based and Airborne Instrumentation for Astronomy
  III, 77350F

\bibitem[{{Rauer} {et~al.}(2014){Rauer}, {Catala}, {Aerts}, {Appourchaux},
  {Benz}, {Brandeker}, {Christensen-Dalsgaard}, {Deleuil}, {Gizon}, {Goupil},
  {G{\"u}del}, {Janot-Pacheco}, {Mas-Hesse}, {Pagano}, {Piotto}, {Pollacco},
  {Santos}, {Smith}, {Su{\'a}rez}, {Szab{\'o}}, {Udry}, {Adibekyan}, {Alibert},
  {Almenara}, {Amaro-Seoane}, {Eiff}, {Asplund}, {Antonello}, {Barnes},
  {Baudin}, {Belkacem}, {Bergemann}, {Bihain}, {Birch}, {Bonfils}, {Boisse},
  {Bonomo}, {Borsa}, {Brand{\~a}o}, {Brocato}, {Brun}, {Burleigh}, {Burston},
  {Cabrera}, {Cassisi}, {Chaplin}, {Charpinet}, {Chiappini}, {Church},
  {Csizmadia}, {Cunha}, {Damasso}, {Davies}, {Deeg}, {D{\'{\i}}az}, {Dreizler},
  {Dreyer}, {Eggenberger}, {Ehrenreich}, {Eigm{\"u}ller}, {Erikson}, {Farmer},
  {Feltzing}, {de Oliveira Fialho}, {Figueira}, {Forveille}, {Fridlund},
  {Garc{\'{\i}}a}, {Giommi}, {Giuffrida}, {Godolt}, {Gomes da Silva},
  {Granzer}, {Grenfell}, {Grotsch-Noels}, {G{\"u}nther}, {Haswell}, {Hatzes},
  {H{\'e}brard}, {Hekker}, {Helled}, {Heng}, {Jenkins}, {Johansen},
  {Khodachenko}, {Kislyakova}, {Kley}, {Kolb}, {Krivova}, {Kupka}, {Lammer},
  {Lanza}, {Lebreton}, {Magrin}, {Marcos-Arenal}, {Marrese}, {Marques},
  {Martins}, {Mathis}, {Mathur}, {Messina}, {Miglio}, {Montalban}, {Montalto},
  {Monteiro}, {Moradi}, {Moravveji}, {Mordasini}, {Morel}, {Mortier},
  {Nascimbeni}, {Nelson}, {Nielsen}, {Noack}, {Norton}, {Ofir}, {Oshagh},
  {Ouazzani}, {P{\'a}pics}, {Parro}, {Petit}, {Plez}, {Poretti}, {Quirrenbach},
  {Ragazzoni}, {Raimondo}, {Rainer}, {Reese}, {Redmer}, {Reffert},
  {Rojas-Ayala}, {Roxburgh}, {Salmon}, {Santerne}, {Schneider}, {Schou},
  {Schuh}, {Schunker}, {Silva-Valio}, {Silvotti}, {Skillen}, {Snellen}, {Sohl},
  {Sousa}, {Sozzetti}, {Stello}, {Strassmeier}, {{\v S}vanda}, {Szab{\'o}},
  {Tkachenko}, {Valencia}, {Van Grootel}, {Vauclair}, {Ventura}, {Wagner},
  {Walton}, {Weingrill}, {Werner}, {Wheatley}, \& {Zwintz}}]{PLATO}
{Rauer}, H., {Catala}, C., {Aerts}, C., {et~al.} 2014, Experimental Astronomy,
  38, 249

\bibitem[{{Reiners} {et~al.}(2010){Reiners}, {Bean}, {Huber}, {Dreizler},
  {Seifahrt}, \& {Czesla}}]{mdwarf}
{Reiners}, A., {Bean}, J.~L., {Huber}, K.~F., {et~al.} 2010, \apj, 710, 432

\bibitem[{{Ricker} {et~al.}(2014){Ricker}, {Winn}, {Vanderspek}, {Latham},
  {Bakos}, {Bean}, {Berta-Thompson}, {Brown}, {Buchhave}, {Butler}, {Butler},
  {Chaplin}, {Charbonneau}, {Christensen-Dalsgaard}, {Clampin}, {Deming},
  {Doty}, {De Lee}, {Dressing}, {Dunham}, {Endl}, {Fressin}, {Ge}, {Henning},
  {Holman}, {Howard}, {Ida}, {Jenkins}, {Jernigan}, {Johnson}, {Kaltenegger},
  {Kawai}, {Kjeldsen}, {Laughlin}, {Levine}, {Lin}, {Lissauer}, {MacQueen},
  {Marcy}, {McCullough}, {Morton}, {Narita}, {Paegert}, {Palle}, {Pepe},
  {Pepper}, {Quirrenbach}, {Rinehart}, {Sasselov}, {Sato}, {Seager},
  {Sozzetti}, {Stassun}, {Sullivan}, {Szentgyorgyi}, {Torres}, {Udry}, \&
  {Villasenor}}]{TESS}
{Ricker}, G.~R., {Winn}, J.~N., {Vanderspek}, R., {et~al.} 2014, in \procspie,
  Vol. 9143, Space Telescopes and Instrumentation 2014: Optical, Infrared, and
  Millimeter Wave, 914320

\bibitem[{Schwab {et~al.}(2016)Schwab, Rakich, Gong, Mahadevan, Halverson, Roy,
  Terrien, Robertson, Hearty, Levi, Monson, Wright, McElwain, Bender, Blake,
  Stürmer, Gurevich, Chakraborty, \& Ramsey}]{NEID}
Schwab, C., Rakich, A., Gong, Q., {et~al.} 2016, Design of NEID, an extreme
  precision Doppler spectrograph for WIYN, , , doi:10.1117/12.2234411

\bibitem[{{Stefansson} {et~al.}(2016){Stefansson}, {Hearty}, {Robertson},
  {Mahadevan}, {Anderson}, {Levi}, {Bender}, {Nelson}, {Monson}, {Blank},
  {Halverson}, {Henderson}, {Ramsey}, {Roy}, {Schwab}, \& {Terrien}}]{NEIDECS}
{Stefansson}, G., {Hearty}, F., {Robertson}, P., {et~al.} 2016, \apj, 833, 175

\bibitem[{{Swift} {et~al.}(2015){Swift}, {Bottom}, {Johnson}, {Wright},
  {McCrady}, {Wittenmyer}, {Plavchan}, {Riddle}, {Muirhead}, {Herzig}, {Myles},
  {Blake}, {Eastman}, {Beatty}, {Barnes}, {Gibson}, {Lin}, {Zhao}, {Gardner},
  {Falco}, {Criswell}, {Nava}, {Robinson}, {Sliski}, {Hedrick}, {Ivarsen},
  {Hjelstrom}, {de Vera}, \& {Szentgyorgyi}}]{MINERVA}
{Swift}, J.~J., {Bottom}, M., {Johnson}, J.~A., {et~al.} 2015, Journal of
  Astronomical Telescopes, Instruments, and Systems, 1, 027002

\end{thebibliography}
%\input{output.aux}
%\input{output.blg}
%\input{main.bbl}

\bibliographystyle{aasjournal}

\end{document}